\documentclass[aps,prb,twocolumn,superscriptaddress]{revtex4}
\usepackage[colorlinks=true,citecolor=blue,linkcolor=blue,breaklinks=true]{hyperref}

\usepackage{color,graphicx}
\usepackage{bm}
\usepackage{amsmath}
\usepackage{amssymb}
\usepackage{times}
\usepackage{physics}
\usepackage{tikz}
\usepackage{standalone}

\definecolor{green}{rgb}{0,0.7,0}
\begin{document}
	
%-----------------------------------------------------%
\newcommand {\beq} {\begin{equation}}
\newcommand {\eeq} {\end{equation}}
\newcommand {\bqa} {\begin{eqnarray}}
\newcommand {\eqa} {\end{eqnarray}}
\newcommand {\ca} {\ensuremath{c^\dagger}}
\newcommand {\ba} {\ensuremath{b^\dagger}}
\newcommand {\Ma} {\ensuremath{M^\dagger}}
\newcommand {\psia} {\ensuremath{\psi^\dagger}}
\newcommand {\fbar} {\ensuremath{\bar{f}}}
\newcommand {\psita} {\ensuremath{\tilde{\psi}^\dagger}}
\newcommand{\lp} {\ensuremath{{\lambda '}}}
\newcommand{\A} {\ensuremath{{\bf A}}}
\newcommand{\Q} {\ensuremath{{\bf Q}}}
\newcommand{\kk} {\ensuremath{{\bf k}}}
\newcommand{\kp} {\ensuremath{{\bf k'}}}
\newcommand{\rr} {\ensuremath{{\bf r}}}
\newcommand{\rp} {\ensuremath{{\bf r'}}}
\newcommand {\ep} {\ensuremath{\epsilon}}
\newcommand{\nbr} {\ensuremath{\langle r r' \rangle}}
\newcommand {\no} {\nonumber}
\newcommand{\up} {\ensuremath{\uparrow}}
\newcommand{\dn} {\ensuremath{\downarrow}}
\newcommand{\rcol} {\textcolor{red}}
\newcommand{\bcol} {\textcolor{blue}}
\newcommand{\lt} {\left}
\newcommand{\rt} {\right}
\newcommand{\comment}[1]{}
\renewcommand{\vec}[1]{{\bf #1}}
\renewcommand{\tilde}{\widetilde}

%-----------------------------------------------------%

%\renewcommand{\vec}[1]{{\bf #1}}

\title{Edge-Plasmon-Induced Magnetization in Higher Chern Insulators}

\author{Prathyush P. Poduval}
\email{ppoduval@uci.edu}

\affiliation{Donald Bren School of Information and Computer Science, University of California, Irvine, CA 92697}

\author{Rofii}
\affiliation{
Research Center for Quantum Physics, National Research and Innovation Agency, Tangerang Selatan 15314, Indonesia.
}
\affiliation{
Department of Physics, Universitas Airlangga, Surabaya 60115, Indonesia
}%

\author{Muhammad S. Ukhtary}
\affiliation{
Research Center for Quantum Physics, National Research and Innovation Agency, Tangerang Selatan 15314, Indonesia.
}
\author{Thomas L. Schmidt}
\affiliation{
Department of Physics and Materials Science, University of Luxembourg, Luxembourg 1511, Luxembourg
}
\author{Eddwi H. Hasdeo}
\email{eddw001@brin.go.id}
\affiliation{
Research Center for Quantum Physics, National Research and Innovation Agency, Tangerang Selatan 15314, Indonesia.
}

\begin{abstract}

Higher Chern insulators host multiple dissipationless chiral edge states, one per unit of Chern number. We study the collective plasmon modes of these edge states at a domain wall between Chern numbers $\mathcal{C}$ and $-\mathcal{C}$, using a semiclassical model with Coulomb interactions. The plasmon dispersion is linear at long wavelengths, with a velocity that increases linearly with $\mathcal{C}$. As in conventional two-dimensional metals, the edge plasmon generates an elliptically polarized electric field that decays into the bulk and drives electrons in circular orbits, producing a static orbital magnetization (inverse Faraday effect). For topological edge plasmons, the magnetization has a nontrivial spatial profile set by two competing length scales: the finite extent of the edge states and the plasmon wavelength. At fixed edge-charge amplitude, it grows strongly with $\mathcal{C}$. At the domain wall, it approaches a finite plateau at low frequency and decays as $1/\omega^2$ at high frequency. Its temperature dependence is nonmonotonic, with a maximum at an intermediate temperature. These results show how the Chern number controls plasmon-induced magnetization, in contrast to conventional edge plasmons.
\end{abstract}

\maketitle

\section{Introduction}

Topological states of matter have become one of the most important breakthroughs in condensed matter physics. The field highlights the crucial role played by the topology of the wave function in dictating the electron dynamics. For example, the quantization of the Hall conductance is closely related to the topological invariant called the Chern number, which is calculated via the integral of the geometric Berry curvature in momentum space. Through the bulk-boundary correspondence, the Chern number is also the number of dissipationless chiral edge modes, which is the origin of the quantization of the Hall conductance. Chern insulators can be found in many materials, including but not limited to magnetically doped three-dimensional (3D) topological insulators~\cite{tcakaev2020comparing,wray2011topological,zhang2009topological,hasan2010colloquium,chang2013experimental,chang2015high}, Hall crystals~\cite{liangfu2025} or even intrinsic materials like $\text{MnBi}_2\text{Te}_4$~\cite{gong2019experimental} and $\rm{LaN_2}$~\cite{lan2}. These materials have been applied in cutting-edge technology such as proximity induced topological superconductivity hosting Majorana zero energy modes~\cite{qi2011topological,tokura2019magnetic}, current induced magnetic controls~\cite{mogi2021current,fujimura2021current}, and spin induced photogalvanic effects~\cite{tokura2019magnetic}.

%Interestingly, wave functions in Chern insulators can be manipulated to allow multiple windings in parametric space resulting in higher Chern number. Recently   higher Chern insulators manifest in ... and tunable up to 5 by stacking multilayers[experimental papers] Higher Chern insulators attract interests because of the possibility to realize even denominator fractional excitation without magnetic field [Neupert flat bands, Kai Sun & Das Sarma flat bands, higher Chern (hC) flat bands ]. Multiple edge channels in hC insulators might be interesting for future opto-electronics [explain]

Interestingly, the wave function of a Chern insulator can be manipulated so that it winds around multiple times in parameter space. This winding would result in a material with higher Chern number, which is called the higher Chern insulator. Recently, a higher Chern insulator was observed by stacking multiple layers of alternating magnetic and undoped topological insulators, resulting in tunable Chern number up to $\mathcal C=5$ \cite{zhaotuning}. There have also been theoretical models about stacking multiple $\mathcal C=1$ Chern insulators to get higher Chern number with a tunable flat band \cite{yang2012topological,zhou2015topological}. Due to the presence of flat bands with non-trivial topology, such models can lead to a realization of fractional excitation states even without the presence of a magnetic field \cite{ge2020high,chang2015zero,skurativska2021flat,sun2011nearly}. Moreover, the multiple dissipationless edge channels in higher Chern insulators without any backscattering will be the key in the development of energy efficient topological electronic devices \cite{gilbert2021topological,zhaotuning,zhou2015topological}.

In this paper, we investigate the collective behavior of chiral edge states in higher Chern insulators ($\mathcal C\ge 1$). We introduce a simple two-band model which can realize any integer Chern number and study the multiple chiral edge states present in it. This model is inspired by the effective two band model for multilayer graphene with a gap~\cite{minlayergraphene}. We  construct a lattice Hamiltonian that can take any arbitrary integer $\mathcal{C}$ as the Chern number and hosts $\mathcal{C}$ chiral edge channels. In the presence of Coulomb interactions, we study how the collective modes of these edge states give rise to charge oscillations propagating along the edge. Within a semi-classical framework, we find that the self generated potential from these oscillations has a decay profile on the axis perpendicular to the edge, towards the bulk. This potential profile generates an elliptically polarized electric field which drives the electrons inside the bulk in a circular fashion, resulting in an induced bulk static orbital moment, known as the inverse Faraday effect \cite{bliokh2017optical,hertel2006theory,fetter1985edge}.

However, compared to the conventional edge plasmons resulting from the non-topological interactions, the higher Chern number edge plasmons (hC plasmons) display stark differences. The dispersion of hC plasmons at long wavelengths is linear with $\omega= v_\mathcal C q$ (where $v_\mathcal C$ is the velocity of the edge plasmon that is proportional to the Chern number) while for conventional 1D edge plasmons the dispersion is $\omega\sim \sqrt{q}$. The dispersion of hC plasmons is tunable by the Chern number, dielectric constant and band gap. Moreover, the frequency dependence of the magnetization induced by hC plasmons differs qualitatively from that of conventional edge plasmons. The edge plasmon can be excited by scanning near-field optical microscopy (SNOM) with a metallic tip \cite{koppens2011graphene,fei2015edge} or by injecting AC current through metal contacts \cite{kumada2013plasmon,kumada2014plasmon}. Recently, the application of a magnetic tip allows the printing of domain wall and control of the chirality of Chern insulators from $+C$ to $-C$ \cite{tokura2019magnetic,nomura2011surface,upadhyaya2016domain}.

\begin{figure*}
\centering
\includegraphics[width=\textwidth]{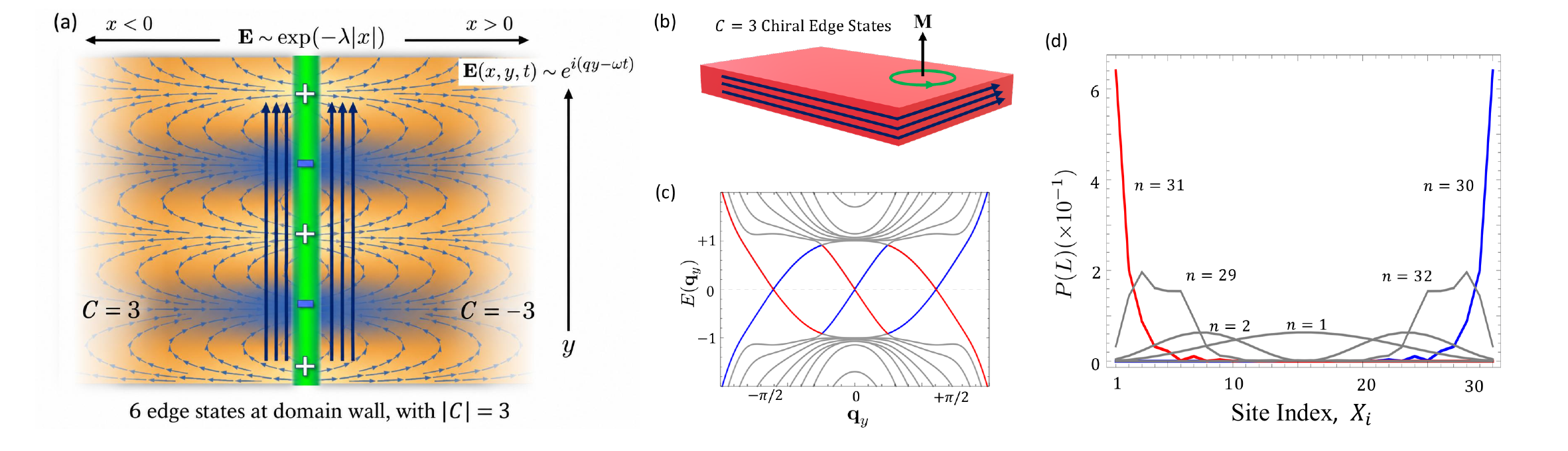}
\caption{Summary of results. (a) Chiral electrons exist at the domain wall separating $C$ and $-C$ higher Chern insulators. The collective mode of these electrons gives rise to a potential profile that oscillates along the domain wall and decays into the bulk $\phi\propto e^{i(qy-\omega t)-\lambda|x|}$ resulting in circularly polarized electric fields ($\vec E=-\nabla \phi$) in the bulk (shown by arrows). The contours indicate equipotentials. The signs along the edge show the oscillating charge density. (b) Schematic of the edge states (black) along the edges, which cause the electrons in the bulk to rotate giving rise to a static magnetization via the inverse Faraday effect. (c) The spectrum of the Hamiltonian with $\mathcal{C}=3$ with a finite lattice size in $x$ direction and in momentum space along $y$ direction. (d) The numerical results of edge states wave function for $L=30$ lattice size along the $x$ direction, with $p_ya/\hbar=0.1$. We see that the middle eigenstates $(n=30, 31)$ are localized on opposite edges. As we consider eigenstates away from the middle of the spectrum, they get more delocalized into the bulk as we move away from the gap.  }
\label{fig1}
\end{figure*}

Our work is organized as follows: In Section~\ref{sec:hamiltonian}, we introduce the Hamiltonian we work with both the continuum and the lattice version. We investigate the spectrum of the edge states and the localization of the eigenstates in a finite sized system with open boundary conditions. In Section~\ref{sec:edgeplasmon}, we discuss our framework for describing the collective excitation of edge plasmons, and how a non-unity Chern number affects the plasmon dispersion and the induced electric field inside the bulk. Using this, in Section~\ref{sec:magnetization} we discuss how the temporally oscillating electric field can give rise to a static magnetization inside the bulk as a second order effect and discuss its properties as a function of frequency. Finally, in Section~\ref{sec:conclusion} we conclude with a summary of our findings and results.

\section{Hamiltonian for Higher Chern Number Insulators}
\label{sec:hamiltonian}

In this section, we discuss a simple two-band model which realizes bands with Chern number $\mathcal C$, where $\mathcal C>0$ is an integer. This Hamiltonian is inspired by the effective model for multilayer graphene with gap~\cite{minlayergraphene}, and we use this Hamiltonian to calculate the longitudinal conductivity and the bulk density at low temperatures. The Hamiltonian is given by
\begin{align}
    H_C = \Delta \begin{pmatrix}\tilde p^{2C}-1&\pi_{+}^C\\ \pi_{-}^C& 1-\tilde p^{2C}\end{pmatrix},
\end{align}
where $\pi_{\pm} =\tilde p e^{\pm i \theta}= \displaystyle\frac{p_x\pm ip_y}{p_0}$, $\tilde{p}=|\pi_\pm|$, and $p_0,\Delta$ are constants that set the scale of momentum and energy respectively. The Hamiltonian can be recast as $H_C = \vec d_{\vec p}\cdot \boldsymbol\tau,$ where $\vec d_{\vec p} =\Delta (\text{Re }\pi_+^C,\text{Im }\pi_{+}^C,\tilde p^{2C} -1)$, $\vec p = (p_x,p_y)$ and $\boldsymbol{\tau}= (\tau^x,\tau^y,\tau^z)$ is the vector of the Pauli matrices. The additional $\tilde p^{2C}$ term in $\tau_z$ is required to ensure integer values of $C$. The point $(d_{\vec p,x},d_{\vec p,y})$ in parameter space  will wind around the origin $C$ times as we vary $\theta $ from $0$ to $2\pi$, resulting in a Chern number solely determined by the parameter $C$, 
\begin{equation}
\mathcal{C}=\displaystyle \int \frac{d^2\vec{k}}{2\pi}\Omega_{\vec p} = C, \quad \Omega_\vec p= \frac12\frac{\vec d_{\vec p} \cdot\left(\partial_{p_x}\vec d_{\vec p} \times \partial_{p_y}\vec d_{\vec p}\right)}{\epsilon_{\vec p}^3},
\end{equation}
where $\Omega_\vec p$ is the Berry curvature and $\epsilon_\vec p=-|\vec d_\vec p|$, the valence band energy.

%However, the key issue is that the current Hamiltonian is in the continuum, and represents only the low energy sector. Due to the non-compactness of the momentum space, we end up with a half quantized Chern number rather than an integer. Moreover,
Next, we investigate the profile of the edge states in our model, for which we use the following regularized lattice Hamiltonian,

\begin{widetext}
\begin{align}
    H^{\textrm{reg}}_C =  \vec d_{\vec p}^{\textrm{reg}}\cdot \boldsymbol\tau=\Delta\begin{pmatrix}\left(2-\cos \displaystyle\frac{p_xa}{\hbar}-\cos \displaystyle\frac{p_ya}{\hbar}\right)^C-1&\left(\sin\displaystyle\frac{p_xa}{\hbar}+i\sin\displaystyle\frac{p_ya}{\hbar}\right)^C\\\left(\sin\displaystyle\frac{p_xa}{\hbar}-i\sin\displaystyle\frac{p_ya}{\hbar}\right)^C&1-\left(2-\cos\displaystyle \frac{p_xa}{\hbar}-\cos\displaystyle \frac{p_ya}{\hbar}\right)^C\end{pmatrix},
\end{align}
\end{widetext}
where $\vec d_{\vec p}^{\textrm{reg}}=\Delta (\text{Re }\Pi_+^C,~\text{Im }\Pi_{+}^C,~1-(2-\cos \displaystyle\frac{p_xa}{\hbar}-\cos \displaystyle\frac{p_ya}{\hbar})^C)$ and $\Pi_+ =\sin\displaystyle\frac{p_xa}{\hbar}+i\sin\displaystyle\frac{p_ya}{\hbar}$.   % The diagonal term is added so that we generate the appropriate winding of the Chern number to get a quantized Chern number $\mathcal C=C$ for the Hamiltonian $H^{\textrm{reg}}_C$. 
  In Appendix~\ref{app:realspace}, we show the derivation of the corresponding real space Hamiltonian. With this Hamiltonian, we investigate the edge states and study how the plasmons in these states affect the bulk. We find that the charge oscillations along the edge will generate an exponentially decaying electric field into the bulk, as shown in Fig.~\ref{fig1}(a). This circularly polarized electric field will cause the electrons to rotate, resulting in a static magnetic moment, as shown in Fig.~\ref{fig1}(b).

To examine the edge states, we specialize to the case $C=3$, and construct a ribbon geometry which is infinite in the $y$ direction and finite with $L$ sites in the $x$ direction. By keeping the $y$ component in the momentum space, we exactly diagonalize the Hamiltonian to find the spectrum as a function of $p_y$. This is shown in Fig.~\ref{fig1}(c) with $14$ lattice sites along the $x$ direction. We find $6$ zero energy states, where $3$ states move along one edge and the remaining move along the opposite edge. In Fig.~\ref{fig1}(c), blue edge states move along $y$ and red edge states move along the negative $y$ direction. Fig.~\ref{fig1}(d) shows the wave function of different modes at $p_y a/\hbar=0.1$ with lattice size $L=30$. The zero energy modes with blue and red colors are localized in two different edges. The bulk states in gray colors have energies outside the band gap and the wave functions that extend into the bulk. %The $n=1$ state is symmetric along $x$ and is delocalized through the bulk, just like the $n=2$ state. The $n=30,31$ states denote the center of the spectrum and correspond to the two edge states which are localized. The $n=29$ and $n=31$ states are majoritarily concentrated to one of the edges however they will not be localized in the thermodynamic limit. Moreover, the blue and red states are localized on opposite edges. 

\section{Plasmons in Topological Edge States}
\label{sec:edgeplasmon}
\begin{figure}
\centering
\includegraphics[width=0.4\textwidth]{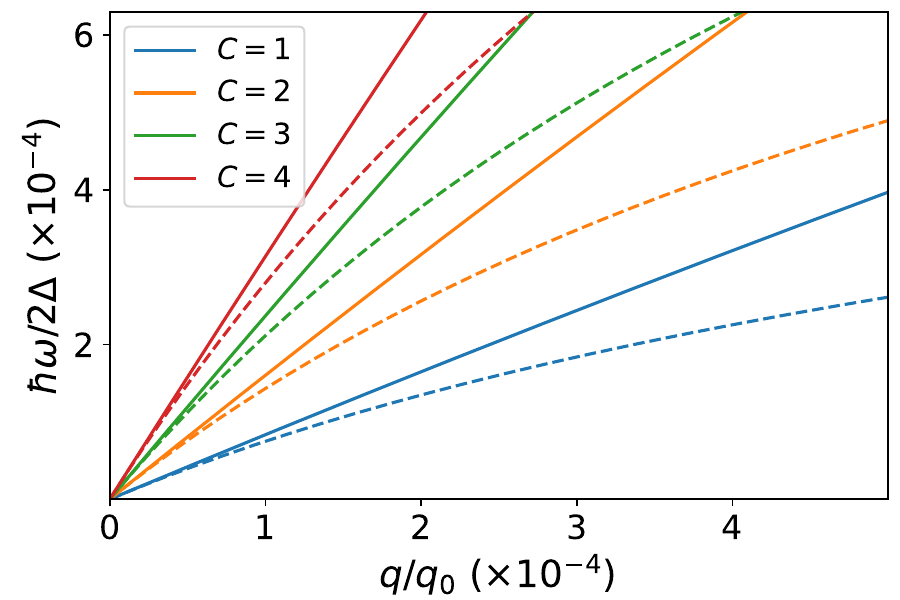}
\caption{Dispersion of edge plasmons: plasmon frequency $\omega$ as a function of wavevector $q$ from Eq.~\eqref{eq:dispersion}. The parameters are $\Delta=0.01~{\rm\ eV}=15~{\rm\ \hbar THz},\  v_0 = 1\ {\rm nm/fs},\  q_0=1/a$ with $a=0.3~{\rm\ nm}$. Solid and dashed lines correspond to $b=10^{-2}q_0$ and $b=10^{-3}q_0$ respectively. The dispersion is linear in $q$, with slope increasing with the Chern number. For a fixed $\omega$, decreasing $b$ increases the corresponding value of $q$. }
\label{fig:dispersion}
\end{figure}

The collective excitation in an electron liquid can be described by the dynamics of charge and current that satisfy the continuity equation, $\partial_t\rho+\nabla\cdot \vec j =0$. In a system with topological edge states, we shall separate the contribution from the edge and from the bulk as $\rho=\rho_b+\rho_e$ and $\vec j=\vec j_b+\vec j_e$.  At finite temperatures, the density in the bulk is given by 
$\rho_b= [e/(2\pi\hbar)^2] \int d^2\vec p[\exp(\epsilon_{\vec{p}}/k_BT)+1]^{-1}\propto g(C) e^{-\Delta/k_BT},$
where we set the Fermi energy $\mu=0$ inside the gap and $g(C)$ is  a nonlinear function of $C$ [see Appendix~\ref{sec:bulkdens}].
The bulk current follows Ohm's law $\vec j_b=\bm{\sigma}(-\nabla \phi)$ in the long wavelength limit, where $\bm{\sigma}$ is the conductivity tensor which comprises a longitudinal component, $\sigma_{xx}= -D/(i\omega)$ with the Drude weight $D\propto C e^{-\Delta/k_B T}$ as indicated by the Drude form and the anomalous Hall component $\sigma_{xy}=-C \sigma_H \tanh \left( \frac{\Delta}{2k_BT}\right)$ with $ \sigma_H=e^2/h$. 

To realistically describe the profile of the edge states, we consider two higher-Chern insulators separated by a boundary, one occupying $x<0$ with Hall conductivity  $\sigma_{xy}=+C\sigma_H$, and the other occupying $x>0$, $\sigma_{xy}=-C\sigma_H$ as shown in Fig.~\ref{fig1}(a). At the boundary, there are $2C$ edge states propagating along the $y$ direction.
The edge current arises from the gapless edge states $\vec j_e=v_0 \rho_e(x,y) \hat{\bf y}$, where  $v_0$ is the group velocity of the edge states.  The edge charge density forms a wave along the edge that decays in the bulk ($x-$ direction) $\rho_e(x,y)=2C \tilde \rho e^{i(q y-\omega t)}e^{-b|x|}$, where  $\tilde \rho$ is the wave amplitude and $b$ is the decay constant. These constitutive equations can be closed by noting that the charge oscillation of free carriers induces the potential profile following the Poisson equation $\nabla^2\phi=-4\pi\delta\rho(\vec r,t)$, where $\delta\rho = \rho -\rho_{b}$ with $\rho_{b}$ being the equilibrium bulk density.

To derive the plasmon dispersion of the edge states, we simplify the calculation by focusing on low temperatures. In this case, the Hall conductivity is quantized $\sigma_{xy}=-C \sigma_H$ and the bulk density $\rho_b$ and thus the longitudinal conductivity $\sigma_{xx}$ are set to zero momentarily [see the expression of plasmon induced bulk magnetization Eq.~\eqref{eq:magnetization} where finite $\rho_b$ is required]. Previous calculations suggest that increasing the temperature (increasing the bulk contribution) will give finite plasmon lifetime but  change the plasmon dispersion  only slightly~\cite{hasdeo2017long}.   

 Two domain walls with different values of Hall conductivity will give
\begin{align}
    \vec{j}_b = \sigma_{yx}(x)(-\partial_x\phi ~\hat{\vec y})+\sigma_{xy}(x)(-\partial_y\phi ~\hat{\vec x}).
\end{align}
To satisfy the continuity equation, the charge density at the edge has to satisfy 
\begin{align} \partial_t\rho_{e}+v_0\partial_y\rho_{e}+(\partial_{x}\sigma_{xy})(-\partial_y\phi(x,y))=0.
\end{align}
Here crucially, the change of $\sigma_{xy}(x)$ across the domain wall will feed into the dynamics of the edge density.
%We can then induce edge plasmons onto the edge current. To describe this, we separate the current and charge density into their bulk and edge parts. This is given by 

The Coulomb interaction generated by the charge oscillations on the edge states is described by the effective one-dimensional Poisson equation in the $x-$direction
\cite{hasdeo2017long,fetter1985edge}. Writing the potential as $\phi(\vec r,t) = 2C\phi_q(x)e^{i(qy-\omega t)}$, we obtain the relation of the potential amplitude with the edge density amplitude as
\begin{align}
    (\partial_x^2-2q^2)\phi_q = -\frac{4\pi}{\kappa} |q|{\tilde{\rho}}e^{-b|x|},\label{eq:poisson}
\end{align}
where $\kappa$ is the dielectric constant; hereafter, we set $\kappa =1$.

We solve Eq.~\eqref{eq:poisson} to find the plasmon potential, 
\begin{align}
    \phi_q(x) = \frac{4\pi \tilde{\rho}}{\sqrt{2}(2q^2-b^2)}  \left({\sqrt{2}q}e^{-b|x|}-be^{-\sqrt{2}|qx|}\right).
\end{align}
At $b=\sqrt{2} q$, the apparent singularity is removed as the integral of Green's function yielding $\phi_q(x) = \frac{\pi\tilde{\rho}}{q} (1+\sqrt{2  }|qx|) e^{-\sqrt{2}|qx|}$. For simplicity, we avoid this resonant condition $b\neq\sqrt{2} q$ and always assume $q>0$.
We obtain the plasmon dispersion [see Appendix~\ref{sec:plas-disp}],
\begin{align}
\label{eq:dispersion}
    \omega=q\left(v_0+\frac{2\sqrt{2}\pi C\sigma_H}{f_q}\right),
\end{align}
where $f_q=1+\sqrt2 |q|/b$ is a form factor correction, accounting for the finite extent $1/b$ of the edge state into the bulk. For $q\ll b$, $f_q=1$ producing a linear spectrum, and the plasmon dispersion speed increases with the Chern number. % For general $q$, $f_q=\frac{1}{v_{\mathcal{C}}}\left(v_0+\frac{\sqrt{2}C\pi\sigma_H}{1+\sqrt{2}|q|/b}\right)$. We choose $\Delta=0.01~{\rm\ eV}$, which corresponds to a gap frequency of about $2\Delta\approx 30~{\rm\ \hbar THz}$. 
In Fig.~\ref{fig:dispersion}, we show the plasmon dispersion focusing primarily on the $30~{\rm\ GHz}$ region, which corresponds to $\sim 50~\mu$m wavelengths for the plasmon.

\section{Magnetization Due to Edge Plasmons}
\label{sec:magnetization}

\begin{figure*}[t]
\centering
\includegraphics[width=0.4\textwidth]{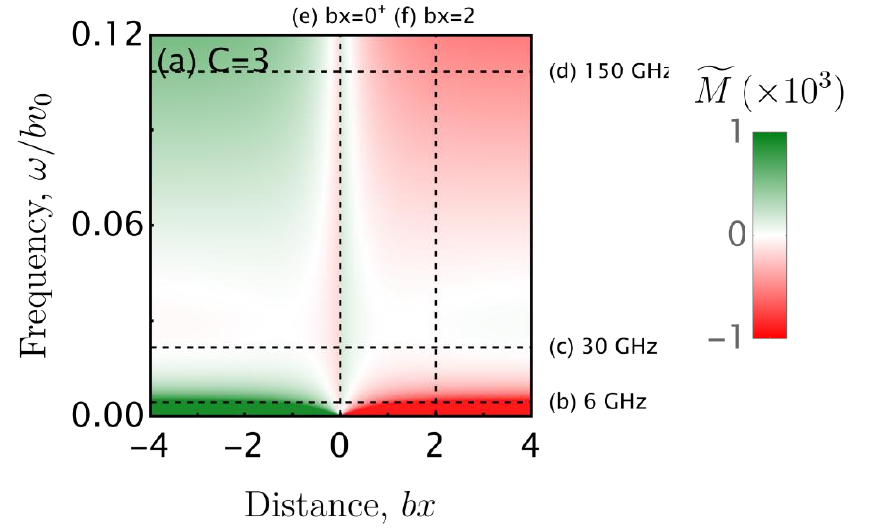}
%\hfill
\includegraphics[width=0.24\textwidth]{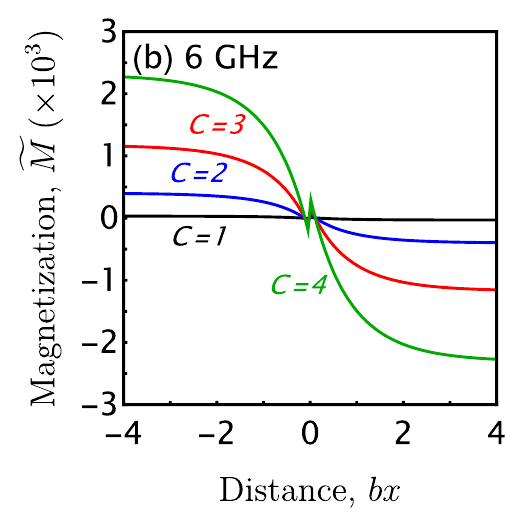}
%\hfill
\includegraphics[width=0.24\textwidth]{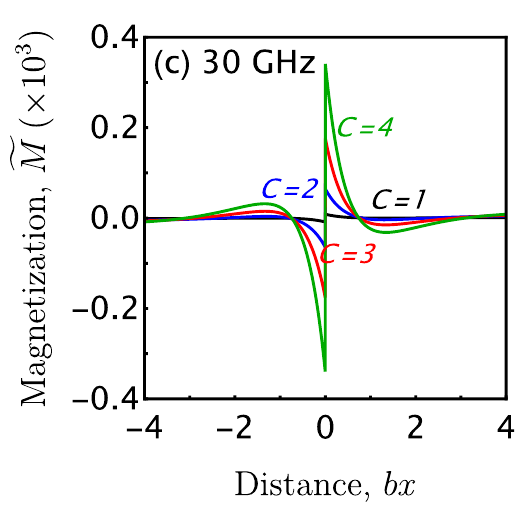}
%\hfill
\includegraphics[width=0.24\textwidth]{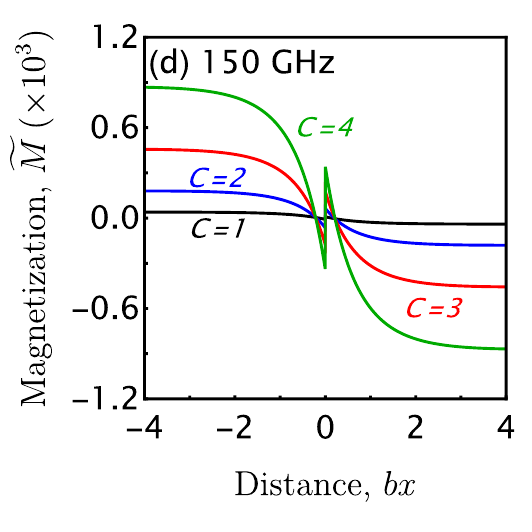}
%\hfill
\includegraphics[width=0.24\textwidth]{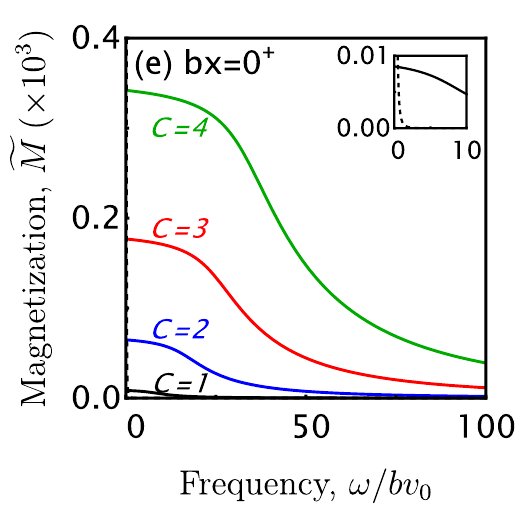}
%\hfill
\includegraphics[width=0.24\textwidth]{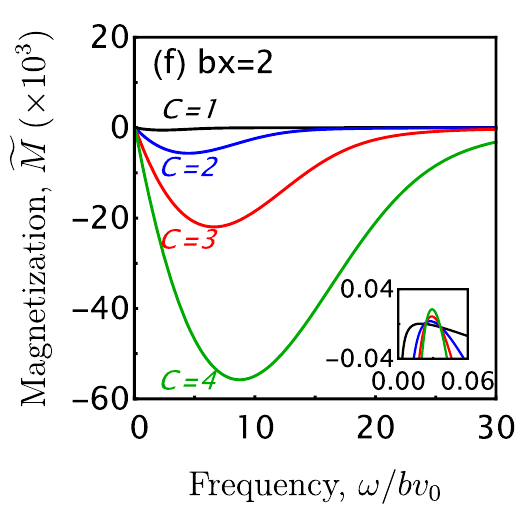}
%\hfill
\includegraphics[width=0.24\textwidth]{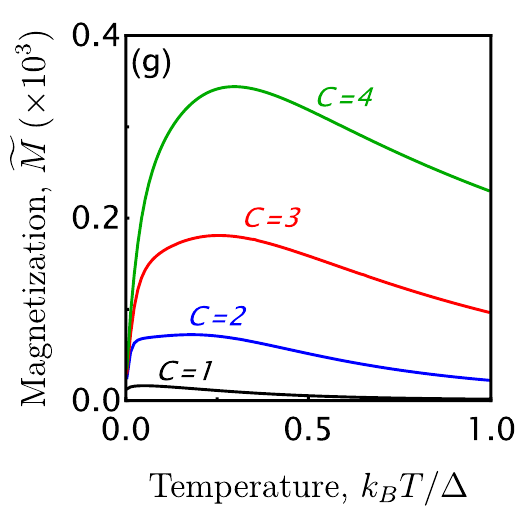}
\caption{(a) 2D map of magnetization $\tilde M= M/M_0$ as functions of distance and frequency for $C=3$ in the low frequency regime $q\ll b$.  (b--d) Magnetization $\tilde M$ as a function of distance from the domain wall for different $C$ values at the frequencies of (b) $6~\mathrm{GHz}$, (c) $30~\mathrm{GHz}$, and (d) $150~\mathrm{GHz}$, respectively. (e-f) Magnetization at (e) $x=0^+$ and (f) $x=2/b$ for different $C$ values as a function of frequency. The dashed black curve in the inset of panel (e) corresponds to the magnetization of a normal metal, $|\tilde{\vec M}_M|$. The insets in panel (e-f) magnify the weak responses on the low-frequency region. The temperature used in panel (a-f) is $T=40~\mathrm{K}$. (g) Temperature dependence of magnetization $\tilde M$ for $\omega=30$ GHz at the domain wall. Here we use $b=10^{-2}/a$ and $v_0=0.75\,\sigma_H$.}
\label{fig:mag0}
\end{figure*}

The edge plasmons generate an electric field that propagates along the edge and decays exponentially into the bulk [see Fig.~\ref{fig1}]. This electric field is elliptically polarized, and electrons move in a circular motion that generates angular momentum, giving rise to the orbital magnetization of the bulk. In this section, we calculate how this magnetization appears as a second order effect in the electric field. %We find that at different frequency regimes, the magnetization shows a distinct dependence behavior on $\omega$, whereas increasing the Chern number in general can enhance the magnetization resulting from topological plasmons.

To calculate the magnetization, we consider the electron to behave like a fluid field. For an electron displaced from its equilibrium position by $\vec d$ and moving with velocity $\vec v$, the mechanical orbital angular momentum is $m_e \vec d \times \vec v$. Therefore, for an equilibrium bulk electron number density $n_b=\rho_b/e$, the corresponding angular-momentum density is $\vec{S}_\textrm{mat}=n_b m_e \vec d \times \vec v$~\cite{bliokh2017optical}. Expressing the physical displacement and velocity in terms of their complex amplitudes gives
%The velocity of this field is related to the electric field by
%\begin{align}
%    \vec v=\frac{[\sigma]\vec E}{\rho_b}
%\end{align}
%Then, the displacement vector of the electron fluid is given by $\partial_t\vec d =\vec v$. If we consider the fourier transforms of $\vec d,\vec v$ and call them $\tilde{\vec d},\tilde{\vec v}$ respectively, then we get that $\tilde{\vec d} = \frac{i}{\omega}\tilde{\vec v}$.
%\end{comment}
\begin{align}
    \vec{S}_\textrm{mat}&=\frac{m_e\rho_b}{4e}(\tilde{\vec{d}}e^{-i\omega t}+\tilde{\vec{d}}^*e^{i\omega t})\times(\tilde{\vec{v}}e^{-i\omega t}+\tilde{\vec{v}}^* e^{i\omega t})\nonumber\\
       & = \frac{m_e}{2\rho_b\omega e}\text{Im }\left( \bm{\sigma}^*\vec{E}^*\times \bm{\sigma}\vec{E}\right) + O (e^{\pm 2 i \omega t}),
\end{align}
where $\tilde {\vec v}=\vec j_b/\rho_b$ is the complex amplitude of the electron-fluid velocity and $\tilde {\vec d} = i\tilde{\vec v}/\omega$ is the corresponding displacement amplitude. The remaining quantities $\vec{E}=-\nabla \phi$ and $\bm{\sigma}$ denote the electric field and conductivity tensor, respectively. 
The expression for angular momentum contains terms that oscillate with frequency $0$ and $\pm 2\omega$. The static term provides a nonzero time-averaged contribution to the static magnetization while the $2\omega$ term will average out to zero. 

The relationship between magnetization and angular momentum is calculated via the gyromagnetic ratio as~\cite{bliokh2017optical,hertel2006theory}

\begin{align}
    \vec{M}=\frac{e}{2m_ec}\vec{S}_{\mathrm{mat}} = \frac{1}{4c\rho_b\omega }\text{Im }\left (\bm{\sigma}^*\vec{E}^*\times \bm{\sigma}\vec{E}\right),\label{eq:magnetization}
\end{align}
where $c$ is the speed of light.
%\textcolor{red}{Mention relation between S and M, change the decay factor for Eq 11, show individual contributions of $\sigma_{xx}$ and $\sigma_{xy}$}
%\begin{align}
%    \vec S_{0}= \frac{-im_e}{2\rho_b\omega e}[\sigma]^*\vec{E}^*\times [\sigma]\vec{E}
%\end{align}
%Using the electronic gyromagnetic ratio, we find the %magnetization to be
%\begin{align}
 %   \vec{M}= \frac{-i}{4\rho_b\omega c}([\sigma]^*\vec{E}^*\times %[\sigma]\vec{E})
%\end{align}
Expanding the expression for the cross product, we obtain the magnetization  as $ \vec{M}=\sum _{i=1}^3 M_i e^{-\lambda_i |x|}\hat{\vec z}$ with three decay constants, namely $\lambda_1= 2\sqrt 2q$, $\lambda_2 =2b $, and $\lambda_3= b+\sqrt 2 q$. These decay constants originate from two characteristic spatial scales: the finite extent of the edge states, characterized by \(1/b\), and the plasmonic length scale, characterized by \(1/q\). The corresponding magnetization amplitudes are
\begin{align}
\tilde{M}_1
&=
\frac{16C^2 e b^3 v_0 \tilde q^2}{\rho_b\omega\left(2\tilde q^2-1\right)^2}
\left[
3\tilde\sigma^t
-
\sqrt{2}s_x\tilde\sigma^2
\right],\\
\tilde{M}_2
&=
\frac{16C^2 e b^3 v_0 \tilde q^2}{\rho_b\omega\left(2\tilde q^2-1\right)^2}
\left[
2(1+\tilde q^2)\tilde\sigma^t
-
2\tilde qs_x\tilde\sigma^2
\right],\\
\tilde{M}_3
&=
\frac{16C^2  e b^3 v_0 \tilde q^2}{\rho_b\omega\left(2\tilde q^2-1\right)^2}
\left[
-
2(2+\sqrt{2}\tilde q)\tilde\sigma^t
+(\sqrt{2}+2\tilde q)s_x\tilde\sigma^2
\right].
\end{align}
where $\tilde q = q/b$, $s_x={\rm sign} (x)$, $\tilde\sigma_{xx}=\sigma_{xx}/\sigma_H$, $\tilde\sigma_{xy}=\sigma_{xy}/\sigma_H$, $\tilde\sigma^2=|\tilde\sigma_{xx}|^2+\tilde\sigma_{xy}^2$,  $\tilde\sigma^t=|\tilde\sigma_{xx}|\tilde\sigma_{xy}$ and 
\beq
M_0
\equiv
\frac{
\pi^2|\tilde{\rho}|^2 \sigma^2_H}{c e b^3 v_0}
\eeq
is a scaling of magnetization, which satisfies $\tilde{\mathbf{M}}=\mathbf{M}/M_0$.
The sign function, ${\rm sign}(x)$, indicates antisymmetric behavior of magnetization due to potential profile $\phi(\vec r,t) \propto e^{-b|x|}$ creating antisymmetric electric field along $x$ direction~[see Fig.~\ref{fig:mag0}(a)]. $\sigma_{xy}$ is implicitly antisymmetric in our domain wall configuration which takes the form $\sigma_{xy} (x) = -C \sigma_H \tanh \left( \frac{\Delta}{2k_BT}\right) {\rm sign} (x)$. 

The shape of magnetization $\tilde M$ in Eq.~\eqref{eq:magnetization} is governed by the competition among the three exponential components and their amplitudes. As shown in Fig.~\ref{fig:mag0}(a), they control the distance and frequency dependence of $\tilde M$. In the low frequency regime $q\ll b$, the $\lambda_1$ or $q$ term has the smallest decay constant and therefore controls the exponential decay far from the domain wall. At this distance, the response is mainly determined by $\tilde M_1$ as shown in Fig.~\ref{fig:mag0}(b-d). However, at a certain frequency, the cancellation between the terms in $\tilde M_1$ strongly suppresses the response far from the domain wall, leaving only significant magnetization in the region near the domain wall, where $\tilde M_2$ and $\tilde M_3$ still have a considerable contribution~[see Fig.~\ref{fig:mag0}(c)]. Interestingly, at specific positions the magnetization crosses zero, signaling a critical point where the electric field becomes linearly polarized before its helicity reverses [see Fig.~\ref{fig:mag0}(b-d)]. 

 Crucially, at $x=0$, the complicated shape of magnetization simplifies to

\beq
\tilde{\mathbf{M}}(0^+)
=
\frac{16C^2 e b^3 v_0}
{\rho_b\omega}
\frac{\tilde q^{\,2}}
{\left(1+\sqrt{2}\tilde q\right)^2}
\tilde\sigma^t \hat{\mathbf z}.
\label{eq:Mag0}
\eeq
where $|\sigma_{xx}|=D/\omega$. The sign change across the domain wall contributes to the jump of magnetization at the boundary as large as $|2M(x=0)|$.
 At low frequency regime $q\ll b$, the factor $\left(\sqrt 2\tilde q +1 \right)^{-2}$ becomes constant. Using plasmon dispersion which is nearly linear, $q\propto \omega/v_\mathcal C$,  the magnetization approaches a finite plateau, which scales approximately as $C^2$ for a fixed $\tilde\rho$, without any divergence at $\omega=0$ as can be seen in Fig.~\ref{fig:mag0}(e). In the large frequency limit $q\gg b$, the asymptotic scaling $M(0^+)\propto C^4/\omega^2$.
At a nonzero distance from the domain wall, the frequency dependence of $\tilde M$ evolves with a divergence in the limit $\omega \to 0^+$, and a sign-reversal within a narrow low-frequency interval. At higher frequency, the magnitude of $\tilde M$ evolves with a peak at a certain frequency which depends on $C$ [see Fig.~\ref{fig:mag0}(f)].

 At a fixed frequency at the domain wall, the temperature dependence enters mainly through the ratio $D/\rho_b$ and through the anomalous Hall conductivity $\sigma_{xy}$. In the near-zero temperature limit, both $D$ and $\rho_b$ vanish, but their ratio approaches a finite value [see Appendix~\ref{sec:bulkdens}]. Consequently, the magnetization also approaches a finite value at low-temperature. As temperature increases, the variation of $D/\rho_b$ initially enhances the magnetization. However, at higher temperature, when $k_B T$ becomes comparable to $\Delta$, the $\sigma_{xy}$ is reduced through the factor
$\tanh(\Delta/2k_BT)$. The competition between these effects produces the nonmonotonic temperature dependence with a maximum at an intermediate temperature as shown in Fig.~\ref{fig:mag0}(g). For larger Chern numbers, this maximum tends to shift toward higher temperature.

We compare the magnetization induced by topological edge states with that of a normal metal. For a normal metal, solve the Poisson equation Eq.~\eqref{eq:poisson} in a half plane geometry with a constant density at $x<0$ and zero density at $x>0$~\cite{fetter1985edge}. As a result, we obtain the magnetization as
\beq
\tilde{\mathbf{M}}_{M}=
-
\frac{\sqrt{2}|\tilde \sigma_{xx}|^2 e b^3 v_0}{\rho_b\omega}
e^{-\sqrt{2}q|x|}
\left(
1-\frac{1}{2}e^{-\sqrt{2}q|x|}
\right)
\hat{\mathbf{z}} .\label{eq:m_norm}
\eeq
The magnetization diverges by $\omega^{-3}$ at zero frequency but becomes negligible quickly at higher frequency [see inset of Fig.~\ref{fig:mag0}(e)]. 

\section{Conclusion}
\label{sec:conclusion}
In this work, we study Chern insulators hosting multiple chiral edge states. By regularizing the continuum Hamiltonian for multilayer graphene, we arrive at a lattice Hamiltonian that hosts an arbitrary number of edge states as described by the Chern number. Using the corresponding tight binding model, we constructed the edge-localized eigenstates, and obtained the corresponding spectrum which shows the presence of zero energy modes. We then consider the effects of plasmons generated by charge oscillations in these edge states. We find the plasmon dispersion is nearly linear in $q$ with $\omega=v_\mathcal C q$, as opposed to normal metals where $\omega\sim \sqrt q$. The plasmon speed $v_\mathcal C$ is found to be linearly increasing with the Chern number. Moreover, the plasmons induce an elliptically polarized electric field that decays exponentially in the direction of the bulk of the material. Such a field produces a static magnetization in the bulk, which is the well known inverse Faraday effect. This magnetization shows an interesting behavior as a function of frequency due to nonzero Hall conductivity. While for large $\omega$, the magnetization at the domain wall is proportional to $1/\omega^2$ , for small frequencies we observe that it becomes approximately independent of $\omega$. Due to this, we find a giant enhancement of magnetization for high frequency regimes in topological materials as opposed to normal metals. This enhancement also increases with the Chern number. Additionally, the magnetization at the domain wall approaches a finite low-temperature plateau and exhibits a nonmonotonic temperature dependence  at finite temperatures.

%The magnetization splits the spin degeneracy of the energy levels and causes an imbalance in the spin populations. This imbalance then diffuses into the bulk via the spin diffusion equation, characterised by the spin diffusion length scale $\lambda_s\sim 10~{\rm \mu m}$ and persists deeper than the static magnetization inside the bulk. 

\begin{acknowledgments}
TLS acknowledges financial support from the National Research Fund Luxembourg under Grants  PRIDE19/14063202/ACTIVE
\end{acknowledgments}

%\bibstyle{aip}
\bibliography{refs}

@article{lan2,
    author = {Li, Linyang and Kong, Xiangru and Chen, Xin and Li, Jia and Sanyal, Biplab and Peeters, François M.},
    title = {Monolayer 1T-LaN2: Dirac spin-gapless semiconductor of p-state and Chern insulator with a high Chern number},
    journal = {Applied Physics Letters},
    volume = {117},
    number = {14},
    pages = {143101},
    year = {2020},
    month = {10},
    issn = {0003-6951},
    doi = {10.1063/5.0023531},
    url = {https://doi.org/10.1063/5.0023531},
    eprint = {https://pubs.aip.org/aip/apl/article-pdf/doi/10.1063/5.0023531/14539661/143101\_1\_online.pdf},
}

@Article{liangfu2025,
author={Paul, Nisarga
and Shavit, Gal
and Fu, Liang},
title={Designing Hall crystals with variable Chern numbers},
journal={Nature Communications},
year={2025},
month={Aug},
day={29},
volume={16},
number={1},
pages={8103},
issn={2041-1723},
doi={10.1038/s41467-025-63320-6},
url={https://doi.org/10.1038/s41467-025-63320-6}
}

@article{tcakaev2020comparing,
  title={Comparing magnetic ground-state properties of the V-and Cr-doped topological insulator (Bi, Sb) 2 Te 3},
  author={Tcakaev, A and Zabolotnyy, VB and Green, RJ and Peixoto, TRF and Stier, F and Dettbarn, M and Schreyeck, S and Winnerlein, M and Vidal, R Crespo and Schatz, S and others},
  journal={Physical Review B},
  volume={101},
  number={4},
  pages={045127},
  year={2020},
  publisher={APS}
}

@article{wray2011topological,
  title={A topological insulator surface under strong Coulomb, magnetic and disorder perturbations},
  author={Wray, L Andrew and Xu, Su-Yang and Xia, Yuqi and Hsieh, David and Fedorov, Alexei V and San Hor, Yew and Cava, Robert J and Bansil, Arun and Lin, Hsin and Hasan, M Zahid},
  journal={Nature Physics},
  volume={7},
  number={1},
  pages={32--37},
  year={2011},
  publisher={Nature Publishing Group}
}

@article{chang2013experimental,
  title={Experimental observation of the quantum anomalous Hall effect in a magnetic topological insulator},
  author={Chang, Cui-Zu and Zhang, Jinsong and Feng, Xiao and Shen, Jie and Zhang, Zuocheng and Guo, Minghua and Li, Kang and Ou, Yunbo and Wei, Pang and Wang, Li-Li and others},
  journal={Science},
  volume={340},
  number={6129},
  pages={167--170},
  year={2013},
  publisher={American Association for the Advancement of Science}
}

@article{chang2015high,
  title={High-precision realization of robust quantum anomalous Hall state in a hard ferromagnetic topological insulator},
  author={Chang, Cui-Zu and Zhao, Weiwei and Kim, Duk Y and Zhang, Haijun and Assaf, Badih A and Heiman, Don and Zhang, Shou-Cheng and Liu, Chaoxing and Chan, Moses HW and Moodera, Jagadeesh S},
  journal={Nature materials},
  volume={14},
  number={5},
  pages={473--477},
  year={2015},
  publisher={Nature Publishing Group}
}

@article{zhang2009topological,
  title={Topological insulators in Bi 2 Se 3, Bi 2 Te 3 and Sb 2 Te 3 with a single Dirac cone on the surface},
  author={Zhang, Haijun and Liu, Chao-Xing and Qi, Xiao-Liang and Dai, Xi and Fang, Zhong and Zhang, Shou-Cheng},
  journal={Nature physics},
  volume={5},
  number={6},
  pages={438--442},
  year={2009},
  publisher={Nature Publishing Group}
}

@article{hasan2010colloquium,
  title={Colloquium: topological insulators},
  author={Hasan, M Zahid and Kane, Charles L},
  journal={Reviews of modern physics},
  volume={82},
  number={4},
  pages={3045},
  year={2010},
  publisher={APS}
}

@article{gong2019experimental,
  title={Experimental realization of an intrinsic magnetic topological insulator},
  author={Gong, Yan and Guo, Jingwen and Li, Jiaheng and Zhu, Kejing and Liao, Menghan and Liu, Xiaozhi and Zhang, Qinghua and Gu, Lin and Tang, Lin and Feng, Xiao and others},
  journal={Chinese Physics Letters},
  volume={36},
  number={7},
  pages={076801},
  year={2019},
  publisher={IOP Publishing}
}

@article{qi2011topological,
  title={Topological insulators and superconductors},
  author={Qi, Xiao-Liang and Zhang, Shou-Cheng},
  journal={Reviews of Modern Physics},
  volume={83},
  number={4},
  pages={1057},
  year={2011},
  publisher={APS}
}

@article{mogi2021current,
  title={Current-induced switching of proximity-induced ferromagnetic surface states in a topological insulator},
  author={Mogi, Masataka and Yasuda, Kenji and Fujimura, Reika and Yoshimi, Ryutaro and Ogawa, Naoki and Tsukazaki, Atsushi and Kawamura, Minoru and Takahashi, Kei S and Kawasaki, Masashi and Tokura, Yoshinori},
  journal={Nature communications},
  volume={12},
  number={1},
  pages={1--6},
  year={2021},
  publisher={Nature Publishing Group}
}

@article{fujimura2021current,
  title={Current-induced magnetization switching at charge-transferred interface between topological insulator (Bi, Sb) 2Te3 and van der Waals ferromagnet Fe3GeTe2},
  author={Fujimura, Reika and Yoshimi, Ryutaro and Mogi, Masataka and Tsukazaki, Atsushi and Kawamura, Minoru and Takahashi, Kei S and Kawasaki, Masashi and Tokura, Yoshinori},
  journal={Applied Physics Letters},
  volume={119},
  number={3},
  pages={032402},
  year={2021},
  publisher={AIP Publishing LLC}
}

@article{tokura2019magnetic,
  title={Magnetic topological insulators},
  author={Tokura, Yoshinori and Yasuda, Kenji and Tsukazaki, Atsushi},
  journal={Nature Reviews Physics},
  volume={1},
  number={2},
  pages={126--143},
  year={2019},
  publisher={Nature Publishing Group}
}

@article{yang2012topological,
  title={Topological flat band models with arbitrary Chern numbers},
  author={Yang, Shuo and Gu, Zheng-Cheng and Sun, Kai and Sarma, S Das},
  journal={Physical Review B},
  volume={86},
  number={24},
  pages={241112},
  year={2012},
  publisher={APS}
}

@article{zhou2015topological,
  title={Topological edge states in single-and multi-layer Bi4Br4},
  author={Zhou, Jin-Jian and Feng, Wanxiang and Liu, Gui-Bin and Yao, Yugui},
  journal={New Journal of Physics},
  volume={17},
  number={1},
  pages={015004},
  year={2015},
  publisher={IOP Publishing}
}

@article{ge2020high,
  title={High-Chern-number and high-temperature quantum Hall effect without Landau levels},
  author={Ge, Jun and Liu, Yanzhao and Li, Jiaheng and Li, Hao and Luo, Tianchuang and Wu, Yang and Xu, Yong and Wang, Jian},
  journal={National Science Review},
  volume={7},
  number={8},
  pages={1280--1287},
  year={2020},
  publisher={Oxford University Press}
}

@article{chang2015zero,
  title={Zero-field dissipationless chiral edge transport and the nature of dissipation in the quantum anomalous Hall state},
  author={Chang, Cui-Zu and Zhao, Weiwei and Kim, Duk Y and Wei, Peng and Jain, Jainendra K and Liu, Chaoxing and Chan, Moses HW and Moodera, Jagadeesh S},
  journal={Physical review letters},
  volume={115},
  number={5},
  pages={057206},
  year={2015},
  publisher={APS}
}

@article{skurativska2021flat,
  title = {Flat bands with fragile topology through superlattice engineering on single-layer graphene},
  author = {Skurativska, Anastasiia and Tsirkin, Stepan S. and Natterer, Fabian D. and Neupert, Titus and Fischer, Mark H.},
  journal = {Phys. Rev. Research},
  volume = {3},
  issue = {3},
  pages = {L032003},
  numpages = {7},
  year = {2021},
  month = {Jul},
  publisher = {American Physical Society},
  doi = {10.1103/PhysRevResearch.3.L032003},
  url = {https://link.aps.org/doi/10.1103/PhysRevResearch.3.L032003}
}

@article{sun2011nearly,
  title={Nearly flatbands with nontrivial topology},
  author={Sun, Kai and Gu, Zhengcheng and Katsura, Hosho and Sarma, S Das},
  journal={Physical review letters},
  volume={106},
  number={23},
  pages={236803},
  year={2011},
  publisher={APS}
}

@article{gilbert2021topological,
  title={Topological electronics},
  author={Gilbert, Matthew J},
  journal={Communications Physics},
  volume={4},
  number={1},
  pages={1--12},
  year={2021},
  publisher={Nature Publishing Group}
}

@inbook{minlayergraphene,
  author = {Hongki Min},
  title = {Electronic properties of multilayer graphene},
  year = {2012},
  month = {2012-03-31},
  publisher = {Graphene Nanoelectronics: Metrology, Synthesis, Properties and Applications, Springer, Berlin, -1},
  url = {https://tsapps.nist.gov/publication/get_pdf.cfm?pub_id=906671},
  language = {en},
}

@article{zhaotuning,
  title={Tuning the Chern number in quantum anomalous Hall insulators},
  author={Zhao, Yi-Fan and Zhang, Ruoxi and Mei, Ruobing and Zhou, Ling-Jie and Yi, Hemian and Zhang, Ya-Qi and Yu, Jiabin and Xiao, Run and Wang, Ke and Samarth, Nitin and others},
  journal={Nature},
  volume={588},
  number={7838},
  pages={419--423},
  year={2020},
  publisher={Nature Publishing Group}
}

@article{fetter1985edge,
  title={Edge magnetoplasmons in a bounded two-dimensional electron fluid},
  author={Fetter, Alexander L},
  journal={Physical Review B},
  volume={32},
  number={12},
  pages={7676},
  year={1985},
  publisher={APS}
}

@article{hertel2006theory,
  title={Theory of the inverse Faraday effect in metals},
  author={Hertel, Riccardo},
  journal={Journal of magnetism and magnetic materials},
  volume={303},
  number={1},
  pages={L1--L4},
  year={2006},
  publisher={Elsevier}
}

@article{koppens2011graphene,
  title={Graphene plasmonics: a platform for strong light--matter interactions},
  author={Koppens, Frank HL and Chang, Darrick E and Garc{\'\i}a de Abajo, F Javier},
  journal={Nano letters},
  volume={11},
  number={8},
  pages={3370--3377},
  year={2011},
  publisher={ACS Publications}
}

@article{fei2015edge,
  title={Edge and surface plasmons in graphene nanoribbons},
  author={Fei, Z and Goldflam, MD and Wu, J-S and Dai, S and Wagner, M and McLeod, AS and Liu, MK and Post, KW and Zhu, S and Janssen, GCAM and others},
  journal={Nano letters},
  volume={15},
  number={12},
  pages={8271--8276},
  year={2015},
  publisher={ACS Publications}
}

@article{nomura2011surface,
  title={Surface-quantized anomalous Hall current and the magnetoelectric effect in magnetically disordered topological insulators},
  author={Nomura, Kentaro and Nagaosa, Naoto},
  journal={Physical review letters},
  volume={106},
  number={16},
  pages={166802},
  year={2011},
  publisher={APS}
}

@article{upadhyaya2016domain,
  title={Domain wall in a quantum anomalous Hall insulator as a magnetoelectric piston},
  author={Upadhyaya, Pramey and Tserkovnyak, Yaroslav},
  journal={Physical Review B},
  volume={94},
  number={2},
  pages={020411},
  year={2016},
  publisher={APS}
}

@article{hasdeo2017long,
  title={Long-lived domain wall plasmons in gapped bilayer graphene},
  author={Hasdeo, Eddwi H and Song, Justin CW},
  journal={Nano letters},
  volume={17},
  number={12},
  pages={7252--7257},
  year={2017},
  publisher={ACS Publications}
}

@article{bliokh2017optical,
  title={Optical momentum and angular momentum in complex media: from the Abraham--Minkowski debate to unusual properties of surface plasmon-polaritons},
  author={Bliokh, Konstantin Y and Bekshaev, Aleksandr Y and Nori, Franco},
  journal={New Journal of Physics},
  volume={19},
  number={12},
  pages={123014},
  year={2017},
  publisher={IOP Publishing}
}

@article{kumada2014plasmon,
  title={Plasmon transport and its guiding in graphene},
  author={Kumada, N and Dubourget, R and Sasaki, K and Tanabe, S and Hibino, H and Kamata, H and Hashisaka, M and Muraki, K and Fujisawa, T},
  journal={New Journal of Physics},
  volume={16},
  number={6},
  pages={063055},
  year={2014},
  publisher={IOP Publishing}
}

@article{kumada2013plasmon,
  title={Plasmon transport in graphene investigated by time-resolved electrical measurements},
  author={Kumada, N and Tanabe, S and Hibino, H and Kamata, H and Hashisaka, M and Muraki, K and Fujisawa, T},
  journal={Nature communications},
  volume={4},
  number={1},
  pages={1363},
  year={2013},
  publisher={Nature Publishing Group UK London}
}

\begin{widetext}

\appendix
\section{The Real Space Hamiltonian}
\label{app:realspace}
Next, to investigate how the edge state looks in hC insulators, we introduce the lattice Hamiltonian,

\begin{align}
    H^{\textrm{reg}}_C =  \vec d_{\vec p}^{\textrm{reg}}\cdot \boldsymbol\tau=\Delta\begin{pmatrix}\left(2-\cos \displaystyle\frac{p_xa}{\hbar}-\cos \displaystyle\frac{p_ya}{\hbar}\right)^C-1&\left(\sin\displaystyle\frac{p_xa}{\hbar}+i\sin\displaystyle\frac{p_ya}{\hbar}\right)^C\\\left(\sin\displaystyle\frac{p_xa}{\hbar}-i\sin\displaystyle\frac{p_ya}{\hbar}\right)^C&1-\left(2-\cos\displaystyle \frac{p_xa}{\hbar}-\cos\displaystyle \frac{p_ya}{\hbar}\right)^C\end{pmatrix},
\end{align}

where $\vec d_{\vec p}^{\textrm{reg}}=\Delta (\text{Re }\Pi_+^C,~\text{Im }\Pi_{+}^C,~1-(2-\cos \displaystyle\frac{p_xa}{\hbar}-\cos \displaystyle\frac{p_ya}{\hbar})^C)$ and $\Pi_+ =\sin\displaystyle\frac{p_xa}{\hbar}+i\sin\displaystyle\frac{p_ya}{\hbar}$.% The diagonal term is added so that we generate the appropriate winding of the Chern number to get a quantized Chern number $\mathcal C=C$ for the Hamiltonian $H^{\textrm{reg}}_C$. 

To generate the appropriate real space lattice Hamiltonian, we write the Hamiltonian as 
\begin{align}
    H^{\textrm{reg}}_C &= \sum_{n_x,n_y=-C}^C \begin{pmatrix}a_{n_x,n_y}&b_{n_x,n_y}\\c_{n_x,n_y}&-a_{n_x,n_y}\end{pmatrix}e^{\displaystyle i\frac{a}{\hbar}\left(n_xp_x+n_yp_y \right)}\\
    &=\sum_{n_x,n_y=-C}^C H_{n_x,n_y}e^{\displaystyle i\frac{a}{\hbar}\left(n_xp_x+n_yp_y \right)}
\end{align}
Note that the summation over $n_x$ and $n_y$ go only up till $C$ and all higher terms will be $0$. Here, $H_{n_x,n_y}$ is a $2\times 2$ block matrix. To find its components, we expand each of the powers using the multinomial expansion to give us

\begin{align}
    (2-\cos f_x - \cos f_y)^C &=(2-e^{if_x}/2-e^{-if_x}/2 -e^{if_y}/2-e^{-if_y}/2)^C\\
    &=\sum_{p+q+r+s+t=C}\frac{C!}{p!q!r!s!t!} 2^{p-q-r-s-t}(-1)^{q+r+s+t}e^{if_x(q-r)+if_y(s-t)}\\
    &=\frac{C!}{2^C}\sum_{p+q+r+s+t=C}\frac{2^{2p}(-1)^{q-r+s-t}}{p!q!r!s!t!} e^{if_x(q-r)+if_y(s-t)},\\
    \hphantom{3}\\
    \hphantom{3}\\
    \hphantom{3}\\
    (\sin f_x+i\sin f_y)^C&=(e^{if_x}/(2i)-e^{-if_x}/(2i) +e^{if_y}/2-e^{-if_y}/2)^C\\
    &=\sum_{q+r+s+t=C}\frac{C!}{q!r!s!t!}\frac{i^{q+r}(-1)^{q+t}}{2^{q+r+s+t}} e^{if_x(q-r)+if_y(s-t)}\\
    &=\frac{C!}{2^C}\sum_{q+r+s+t=C}\frac{i^{q+r}(-1)^{q+t}}{q!r!s!t!} e^{if_x(q-r)+if_y(s-t)},
\end{align}
where $f_x=p_xa/\hbar$ and $f_y=p_ya/\hbar$. Thus, the components of $H_{n_x,n_y}$ are given by 

\begin{align}
    a_{n_x,n_y}&= -\delta_{n_x,0}\delta_{n_y,0}+ \frac{C!}{2^C}\sum_{p,q,r,s,t}\frac{2^{2p}(-1)^{n_x+n_y}}{p!q!r!s!t!}\\
    b_{n_x,n_y}&= \frac{C!}{2^C}\sum_{q,r,s,t}\frac{i^{q+r}(-1)^{q+t}}{q!r!s!t!} \\
    c_{n_x,n_y}&= \frac{C!}{2^C}\sum_{q,r,s,t}\frac{i^{q+r}(-1)^{q+s}}{q!r!s!t!}
\end{align}

\comment{\begin{align}
    a_{n_x,n_y}&= \delta_{n_x,0}\delta_{n_y,0}- \frac{C!}{2^C}\sum_{p,q,r,s,t}\frac{2^{2p}(-1)^{n_x+n_y}}{p!q!r!s!t!}\\
    b_{n_x,n_y}&= \sum_{q,r,s,t}\frac{C!}{q!r!s!t!}(2i)^{-C}(-1)^{r+t}i^{s+t}\\
    c_{n_x,n_y}&= \sum_{q,r,s,t}\frac{C!}{q!r!s!t!}(2i)^{-C}(-1)^{r+t}(-i)^{s+t}
\end{align}}
Where the summation for $a_{n_x,n_y}$ is over all integers $p,q,r,s,t$ that satisfy $p+q+r+s+t=C$, $q-r=n_x$ and $s-t=n_y$. The summation for $b_{n_x,n_y}$ and $c_{n_x,n_y}$ is over all integers $q,r,s,t$ that satisfy $q+r+s+t=C$, $q-r=n_x$ and $s-t=n_y$. These formulas allow us to perform an inverse Fourier transform to construct the real space Hamiltonian whose components are
\begin{align}
    H_C^{reg}(\vec{r}_2,\vec{r}_1) &= H_{x_2-x_1,y_2-y_1}
\end{align}
Where $\vec r_1=(x_1,y_1)$ and $\vec r_2=(x_2,y_2)$. 
\section{Plasmon Dispersion of Topological Edge States \label{sec:plas-disp}}
Here we review the derivation of plasmon dispersion in Eq.~\eqref{eq:dispersion}. We consider the case of a domain wall between two models with Chern number $C$ and $-C$. Thus, there will be $2C$ edge states propagating along the domain wall. Moreover, the edge states have a finite profile along the direction perpendicular to the edge $x$, such that $\delta \rho_{e,q} \sim e^{-b|x|}e^{iqy-i\omega t}$ as an example. We will keep this profile generic for now and write it in terms of the form factor. 
The edge is assumed to be along the $y$ direction, with the domain change from $C$ to $-C$ chern number happening along the $x$ direction.

For simplicity, we  assume that the bulk charge density is $0$. From this, we get the continuity equation at the edge to be 

\begin{align}
    \partial_t\delta\rho_{e,q}+v_0\partial_y\delta\rho_{e,q}+(\partial_{x}\sigma_{xy})(-\partial_y\phi_q(x))=0
\end{align}

$\sigma_{xy}=C\sigma_H t(x)$, where $t(x)$ is a smooth function that changes from $+1$ at $x=0^-$ to $-1$ at $x=0^+$  only across the domain wall in the vicinity of $x=0$, and remains constant otherwise. Thus $\partial_x\sigma_{xy}=-2C\sigma_H\delta(x)$ is nonzero only around $x\sim 0$. This sign is tied to the propagation direction of the edge modes: the Hall current $j_x=\sigma_{xy}E_y$ feeding the wall is $2C\sigma_H E_y$, which is the chiral-anomaly source term for $2C$ modes moving along $+\hat{\vec y}$ (so $v_0>0$). Hence the Hall term in the plasmon speed adds to $v_0$ when the modes co-propagate with the sign of $\sigma_{xy}$ chosen here. Thus we can therefore approximate $\phi_q(x)=\phi_q(x)|_{x=0}$ in the above equation. Integrating the above equation across $x$, we get 
\begin{align}
    (-i\omega+v_0(iq))Q_{e,q}=-2C\sigma_H(iq)\phi_0,\label{eq:cont}
\end{align}
where $\phi_0\equiv \phi_q(x)|_{x=0}$ and $Q_{e,q} = \int dx \rho_{e,q}$ is the charge density at the edge. The relationship between the potential and charge distribution is given by the effective $1D$ Coulomb kernel given by
\begin{align}
    (\partial_x^2-2q^2)\phi_q(x)=-\frac{4\pi}{\kappa} |q| \delta\rho_q(x),
\end{align}
with the solution
\begin{align}
    \phi_q(x) = -\frac{4\pi}{\kappa}|q|C_0 \int dx' G(x-x')\delta\rho_{e,q}(x')=\frac{4\pi}{\kappa}|q|C_0 \int dx' G(x')\delta\rho_{e,q}(x-x')
\end{align}
where $G(x) =e^{-\sqrt{2}|qx|}$. Note that $\delta\rho_e$ is continuous at $x=0$, but its derivative is not. Using this, if we integrate the Coulomb equation about $x=0$, we can find that $\partial_x\phi_q$ is continuous at $x=0$. Using the equation itself and the continuity of $\phi_q(x)$, we find that $\partial_x^2$ is also continuous at $x=0$. Thus, the first discontinuity occurs in the third order derivative given by 
\begin{align}
    \partial_x^3\phi(x)|_{x=0^+}-\partial_x^3\phi(x)|_{x=0^-}=-\frac{4\pi}{\kappa}|q|(\partial_x\delta\rho_{e,q}|_{x=0^+}-\partial_x\delta\rho_{e,q}(x)|_{x=0^-})
\end{align}
Now, we evaluate the difference in the third derivative. \begin{align}
      \partial_x^3\phi(x)|_{x=0^+}-\partial_x^3\phi(x)|_{x=0^-}&=\frac{4\pi}{\kappa}|q|C_0\lim_{\eta\to0} \int dx'G(x')\left( \delta\rho_{e,q}'''(\eta-x')- \delta\rho_{e,q}'''(-\eta-x')\right)\\
      &=\frac{4\pi}{\kappa}|q|C_0\lim_{\eta\to0} \int_{-\eta}^\eta dx'G(x')\left( \delta\rho_{e,q}'''(\eta-x')- \delta\rho_{e,q}'''(-\eta-x')\right)
\end{align} 

From the boundary condition we obtain $C_0=-\frac{1}{2\sqrt{2}|q|}$. Thus, Eq.~\eqref{eq:cont} becomes 
\begin{align}
    \omega - v_0 q =2C\sigma_H q \frac{\phi_0}{Q_{e,q}}
\end{align}

This gives 
\begin{align}
    \omega= q\left(v_0+\frac{2\sqrt{2}\pi C\sigma_H}{\kappa f_q}\right), 
\end{align}
where $f_q= \frac{\sqrt{2}\pi}{\kappa}\frac{Q_{e,q}}{\phi_q(x=0)}=\frac{\sqrt{2}\pi}{\kappa}\frac{\int dx \rho_{e,q}(x) }{\frac{\sqrt{2}\pi}{\kappa}\int dx G(x)\delta\rho_{e,q}(x)}=\frac{\int dx \rho_{e,q}(x) }{\int dx G(x)\delta\rho_{e,q}(x)}$. For $\delta\rho_{e,q}(x) \sim e^{-b|x|}, f_q= 1+\frac{\sqrt{2}|q|}{b}$. Thus, the plasmon dispersion is 

\begin{align}
    \omega= q\left(v_0+\frac{2\sqrt{2}\pi C\sigma_H}{\kappa \left(1+\frac{\sqrt{2}|q|}{b}\right)}\right).
\end{align}

\section{Derivation of Conductivity Matrix}

In this appendix, we derive the conductivity tensor used in the main text. Our goal is to obtain the longitudinal Drude-like response $\sigma_{xx} = -D/(i\omega)$ and the anomalous Hall response $\sigma_{xy}$ from semiclassical Boltzmann transport in the presence of Berry curvature. By assuming the sample is clean and devoid of scattering, the Boltzmann transport equation for the full distribution function $f(\mathbf{p}, \mathbf{r}, t)$ can be written as
\begin{equation}\frac{\partial f}{\partial t}
+ \dot{\mathbf{r}} \cdot \frac{\partial f}{\partial \mathbf{r}}
+ \dot{\mathbf{p}} \cdot \frac{\partial f}{\partial \mathbf{p}} = 0\,,
\label{eq:BEg}
\end{equation}
where $\dot{\mathbf{r}}$ represents the velocity and $\dot{\mathbf{p}}$ represents the force. Substituting
\begin{equation}
    \dot{\vec{r}}=\vec{v}_{\vec p}+ \frac{e\vec E}{\hbar}\times \boldsymbol{\Omega}_{\vec{p}} \,,\qquad \dot{\mathbf{p}}=e\mathbf{E}\,,
\end{equation}
into the Eq.~\eqref{eq:BEg} gives us
\begin{equation}
\frac{\partial f}{\partial t}
+ \left( \vec{v}_{\vec p}+ \frac{e\vec E}{\hbar}\times \boldsymbol{\Omega}_{\vec{p}} \right) \cdot \frac{\partial f}{\partial \mathbf{r}}
+ e\mathbf{E} \cdot \frac{\partial f}{\partial \mathbf{p}} = 0\,,
\end{equation}
where $\vec{v}_{\vec p}=\frac{\partial \epsilon_\vec{p}}{\partial\vec{p}}$ is the band velocity and $\boldsymbol{\Omega}_{\vec{p}}$ is the Berry curvature.

For a potential that exponentially decays into the bulk in the $\hat x$ direction and oscillates in the $\hat y$ direction, the potential function is given by $\phi(r,t) = \phi_0 \exp\left[-\lambda |x| +i(qy-\omega t)\right]$. Specializing the domain $x<0$, the electric field is
\begin{align}
\mathbf{E}&=-\nabla\phi\,,
\\
&=-\phi_0(\lambda ,iq)\exp\left[\lambda x +i(qy-\omega t)\right]\,,
\\
&=\phi_0\, \vec{n}\,\exp\left[\lambda x +i(qy-\omega t)\right]\,,
\label{eq:EF}
\end{align}
where $\vec{n}= (-\lambda,-iq)$. We expand the Fermi distribution function as a power series in $\phi$ as $f=f_0+f_1\phi_0+f_2\phi_0^2+...$ where $f_0$ is the equilibrium Fermi distribution function which is given by 
\begin{equation}
f_0(\vec p) = \frac{1}{\exp \left[ \frac{(\epsilon_\mathbf{p}-\mu) }{k_BT}\right] +1}\,.
\end{equation}
Assuming that the magnitude of the electric field is small, we consider only the linear order response given by $f=f_0+f_1\phi_0$ and we get
\begin{equation}
\frac{\partial (f_0+f_1\phi_0)}{\partial t}
+ \left( \vec{v}_{\vec p}+ \frac{e\vec E}{\hbar}\times \boldsymbol{\Omega}_{\vec{p}} \right) \cdot \frac{\partial (f_0+f_1\phi_0)}{\partial \mathbf{r}}
+ e\mathbf{E} \cdot \frac{\partial (f_0+f_1\phi_0)}{\partial \mathbf{p}} = 0\,.
\end{equation}
Since $f_0$ is the equilibrium distribution, it does not depend explicitly on time and is homogeneous in real space, which gives us
\begin{equation}
\phi_0\frac{\partial f_1}{\partial t}
+ \phi_0\, \vec{v}_{\vec p} \cdot \frac{\partial f_1}{\partial \mathbf{r}}
+ \phi_0\left( \frac{e\vec E}{\hbar}\times \boldsymbol{\Omega}_{\vec{p}}\right) \cdot \frac{\partial f_1}{\partial \mathbf{r}}
+ e\mathbf{E} \cdot \frac{\partial f_0}{\partial \mathbf{p}} + \phi_0\,e\mathbf{E} \cdot \frac{\partial f_1}{\partial \mathbf{p}} = 0\,.
\label{eq:BE2}
\end{equation}
Substituting Eq.~\eqref{eq:EF} into Eq.~\eqref{eq:BE2} and collecting only the term first order in $\phi_0$ in the Boltzmann transport equation gives us
\begin{equation}
\frac{\partial f_1}{\partial t}
+ \vec{v}_{\vec p} \cdot \frac{\partial f_1}{\partial \mathbf{r}}
+ e\,\mathbf{n} \cdot \frac{\partial f_0}{\partial \mathbf{p}}\exp\left[\lambda x +i(qy-\omega t)\right] = 0\,.
\end{equation}

Assuming a trial solution of the form $f_1=A \exp\left[\lambda x +i(qy-\omega t)\right]$ gives us
\begin{equation}
    (-i\omega) A + \vec v_{\vec p}\cdot(-\vec{n})A+e \vec{n}\cdot\frac{\partial f_0}{\partial \mathbf{p}} = 0\,.
\end{equation}
Using $\frac{\partial f_0}{\partial \mathbf{p}}=  {\vec v}_{\vec p} \frac{\partial f_0}{\partial \mathbf{\epsilon_\vec{p}}}$, we get
\begin{equation}
    A =\frac{e(\vec{n}\cdot \vec v_{\vec p})}{i\omega +\vec{n}\cdot \vec v_{\vec p}} \frac{\partial f_0}{\partial \mathbf{\epsilon_\vec{p}}}\,.
\end{equation}
The distribution function becomes 
\begin{equation}
    f
    = f_0 + \frac{e(\vec E\cdot \vec v_{\vec p})}{i\omega +\vec{n}\cdot \vec v_{\vec p}} \frac{\partial f_0}{\partial \mathbf{\epsilon_\vec{p}}}\,. 
\end{equation}
In the long wavelength limit, we can write $i\omega +\vec{n}\cdot \vec v_{\vec p} \approx i\omega$ resulting in
\begin{equation}
    f
    = f_0 + \frac{e(\vec E\cdot \vec v_{\vec p})}{i\omega} \frac{\partial f_0}{\partial \mathbf{\epsilon_\vec{p}}}\,. 
\label{eq:DF}
\end{equation}

Using the distribution function provided by Eq.~\eqref{eq:DF}, and by defining $\rho = e\int \frac{d^2\vec p}{(2\pi\hbar)^2}f_0$, we can obtain the average velocity given by
\begin{align}
    \vec v&= \langle \dot{\vec r}\rangle\,,\\
    &= \frac{\int \frac{d^2\vec p}{(2\pi\hbar)^2} f\,\dot{\vec r}}{\int \frac{d^2\vec p}{(2\pi\hbar)^2}f}\,,\\
    &=\frac{e}{\rho}\int \frac{d^2\vec p}{(2\pi\hbar)^2} f\,\dot{\vec r}\,,\\
    &=\frac{e}{\rho}\int \frac{d^2\vec p}{(2\pi\hbar)^2} \left( f_0 + \frac{e(\vec E\cdot \vec v_{\vec p})}{i\omega} \frac{\partial f_0}{\partial \mathbf{\epsilon_\vec{p}}} \right) \left( \vec{v}_{\vec p} + \frac{e\vec E}{\hbar}\times \boldsymbol{\Omega}_{\vec{p}}\right)\,,\\
    &=\frac{e}{\rho}\int \frac{d^2\vec p}{(2\pi\hbar)^2} \left( f_0\vec{v}_{\vec p} + \vec{v}_{\vec p} \frac{e(\vec E\cdot \vec v_{\vec p})}{i\omega} \frac{\partial f_0}{\partial \mathbf{\epsilon_\vec{p}}} + \frac{e\vec E}{\hbar}\times \boldsymbol{\Omega}_{\vec{p}}f_0+  \frac{e(\vec E\cdot \vec v_{\vec p})}{i\omega} \frac{\partial f_0}{\partial \mathbf{\epsilon_\vec{p}}} \left( \frac{e\vec E}{\hbar}\times \boldsymbol{\Omega}_{\vec{p}}\right) \right)
    \label{eq:avvel}\,,\\
     &=\frac{e}{\rho}\int \frac{d^2\vec p}{(2\pi\hbar)^2} \left( \vec{v}_{\vec p} \frac{e(\vec E\cdot \vec v_{\vec p})}{i\omega} \frac{\partial f_0}{\partial \mathbf{\epsilon_\vec{p}}} + \frac{e\vec E}{\hbar}\times \boldsymbol{\Omega}_{\vec{p}}f_0 \right)\,.
\end{align}
In Eq.~\eqref{eq:avvel}, we can ignore the first term of the integrand because it yields zero after integration and the fourth term of the integrand because it is of second order. The total current is given by 
\begin{align}
    \vec j &= \rho \vec v\,,\\ 
    &= e\int \frac{d^2\vec p}{(2\pi\hbar)^2} \left( \vec{v}_{\vec p} \frac{e(\vec E\cdot \vec v_{\vec p})}{i\omega} \frac{\partial f_0}{\partial \mathbf{\epsilon_\vec{p}}} + \frac{e\vec E}{\hbar}\times \boldsymbol{\Omega}_{\vec{p}}f_0 \right)\,.
\end{align}
Since $j_i=\sigma_{ij}\, E_j$, we can obtain conductivity tensor $\bm{\sigma}$ given by
\begin{equation}
\sigma_{ij}
=e^2
\int \frac{d^2 \vec p}{(2\pi\hbar)^2}
\left(
\frac{1}{i\omega}\, (\vec v_{\vec p})_i (\vec v_{\vec p})_j\, \frac{\partial f_0}{\partial \epsilon_{\vec p}}
+
\epsilon_{ijk}\,\frac{f_0}{\hbar}\,(\vec\Omega_{\vec p})_k \right).
\end{equation}
The longitudinal component and anomalous Hall component can be written as
\begin{equation}
\sigma_{xx}
=\frac{e^2}{i\omega}
\int \frac{d^2 \vec p}{(2\pi\hbar)^2}
(\vec v_{\vec p})^2_x\, \frac{\partial f_0}{\partial \epsilon_{\vec p}},
\label{eq:sigmaxx}
\end{equation}
\begin{equation}
\sigma_{xy}
=\frac{e^2}{\hbar}
\int \frac{d^2 \vec p}{(2\pi\hbar)^2}
(\vec\Omega_{\vec p})_z f_0.
\label{eq:sigmaxy}
\end{equation}
The longitudinal conductivity in Eq.~\eqref{eq:sigmaxx} can be written in the Drude form
\begin{equation}
    \sigma_{xx}(\omega)=-\frac{D}{i\omega},
    \label{eq:sigmaxx_drude}
\end{equation}

\section{Bulk density, Drude weight, and Chern number and temperature dependence \label{sec:bulkdens}}

\begin{figure*}[t]
\centering
\includegraphics[width=0.24\textwidth]{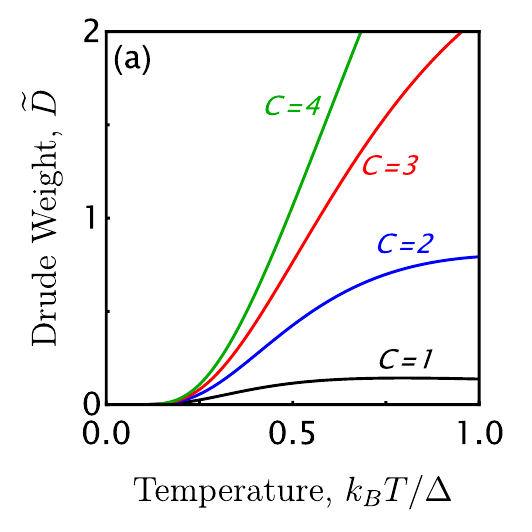}
%\hfill
\includegraphics[width=0.24\textwidth]{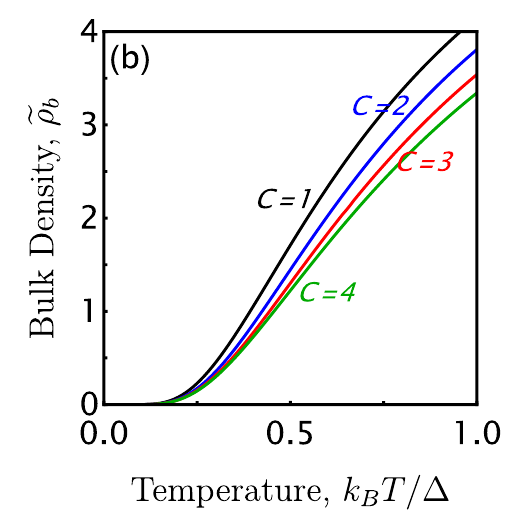}
%\hfill
\includegraphics[width=0.24\textwidth]{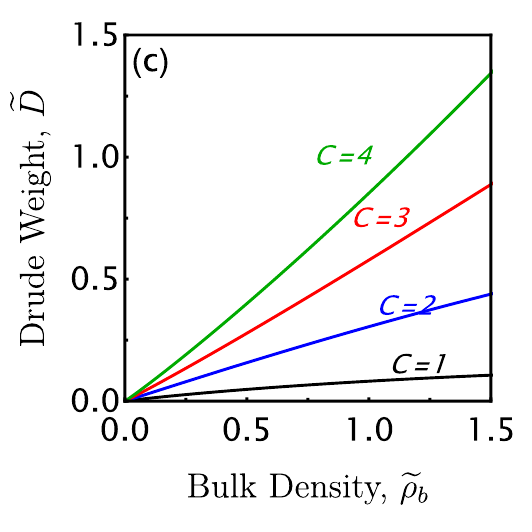}
%\hfill
\includegraphics[width=0.24\textwidth]{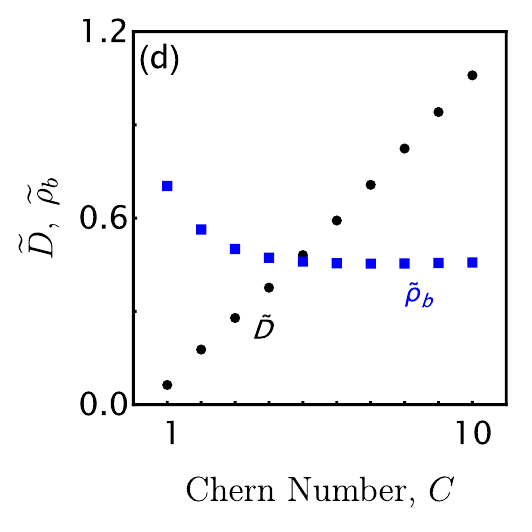}
\caption{(a) Drude weight $(\tilde D=D/D_0)$ and (b) bulk density $(\tilde\rho_b=\rho_b/\rho_{b0})$ as a function of temperature for different Chern numbers. Both quantities vanish in the zero-temperature limit. (c) $\tilde D$ plotted as a function of $\tilde \rho_b$. (d) $\tilde D$ and $\tilde \rho_b$ plotted as a function of $C$ at fixed $T=40~\mathrm{K}$ and $\omega=30~\mathrm{GHz}$. Here $D_0=\frac{e^2 \Delta}{\left(2\pi \hbar\right)^2}$ and $\rho_{b0}=\frac{e}{\left(2\pi a \right)^2}$
}
\label{fig:appendix_drude_density}
\end{figure*}

The magnetization in Eq.~\eqref{eq:Mag0} depends on the longitudinal response through
$|\sigma_{xx}|=D/\omega$ and is proportional to $1/\rho_b$. To obtain the magnetization as shown in Fig.~\ref{fig:mag0}(g), we have to calculate the thermal factor entering the magnetization, which is the ratio $D(T,C)/\rho_b(T,C)$, together with the anomalous Hall conductivity $\sigma_{xy}(T,C)$.  The Drude weight can be defined by comparing Eq.~\eqref{eq:sigmaxx} and Eq.~\eqref{eq:sigmaxx_drude} to get
\begin{equation}
D=-e^2 \int \frac{d^2 \vec p}{(2\pi\hbar)^2}
(\vec v_{\vec p})^2_x\, \frac{\partial f_0}{\partial \epsilon_{\vec p}},
\label{eq:drude_weight_appendix}
\end{equation}
while the bulk density  can be written as

\begin{equation}
    \rho_b= e \int\frac{d^2\vec p}{(2\pi\hbar)^2}\frac{1}{\exp(\epsilon_{\vec{p}}/k_BT)+1}.
\label{eq:bulkdensity_appendix}
\end{equation}
Using these definitions, the Drude weight and the bulk density can be calculated numerically as shown in Fig.~\ref{fig:appendix_drude_density}(a-c). From Fig.~\ref{fig:appendix_drude_density}(a) and Fig.~\ref{fig:appendix_drude_density}(b), it can be seen that increasing temperature
enhances both quantities, but they do not grow identically. Both $D$ and $\rho_b$ vanish in the zero-temperature limit. However, Fig.~\ref{fig:appendix_drude_density}(c) shows that $D$ depends approximately linearly on $\rho_b$ near the origin, indicating that $\lim_{T \to 0}D/\rho_b$ is finite.

The ratio $D/\rho_b$ is also important to determine the Chern number scaling of magnetization, which, by using Eq.~\eqref{eq:Mag0} at fixed temperature and frequency, can be written as
\beq
\vec M(x=0^+)\propto\frac{C^3}{\left(\frac{b}{q(C)} +\sqrt{2}\right)^2} \frac{D(C)}{\rho_b(C)}.
\label{eq:MC_appendix}
\eeq
The Drude weight increases approximately linearly with the Chern number, whereas
$\rho_b$ varies strongly only for the first few Chern numbers and then
becomes weakly dependent on $C$. As a result, $D/\rho_b$ increases
almost linearly with $C$ over the range as shown in Fig.~\ref{fig:appendix_drude_density}(d). Using this approximation, Eq.~\eqref{eq:MC_appendix} becomes
\beq
\vec M(x=0^+)\propto\frac{C^4}{\left(\frac{b}{q(C)} +\sqrt{2}\right)^2}.
\label{eq:MC_appendix2}
\eeq
At high $q$ or $\omega$, the denominator in Eq.~\eqref{eq:MC_appendix2} is just a constant, and this equation reduces to $M(x=0^+)\propto C^4$. In contrast, using plasmon dispersion in Eq.~\eqref{eq:dispersion} at low $q$ or $\omega$, resulting in
\beq
\vec M(x=0^+)\propto\frac{C^4}{\left(v_0+2\sqrt{2}\pi C \sigma_H \right)^2}.
\label{eq:MC_appendix3}
\eeq
Assuming $v_0$ is much smaller than $\sigma_H$, we obtain $M(x=0^+)\propto C^2$.

\end{widetext}

\end{document}